\documentclass[twocolumn,prl,showpacs,a4]{revtex4}
\usepackage{amsmath,amssymb,amsthm}

\newcommand{\ket}[1]{|#1\rangle}

\newcommand{\one}{\mbox{$1 \hspace{-1.0mm}  {\bf l}$}}

\begin{document}
\title{Holonomic quantum computation in the presence of decoherence}
\author{I. Fuentes-Guridi$^{\diamond}$$^\dagger$, F. Girelli$^\dagger$, E. Livine$^\dagger$}
\affiliation{$^{\diamond}$ Centre for Quantum Computation,
Clarendon Laboratory, University of Oxford, Parks Road OX1 3PU
\\$^\dagger$Perimeter Institute, 35 King
Street North Waterloo, Ontario Canada N2J 2W9}

\begin{abstract}
We present a scheme to study non-abelian adiabatic holonomies for
open Markovian systems. As an application of our framework, we
analyze the robustness of holonomic quantum computation against
decoherence. We pinpoint the sources of error that must be
corrected to achieve a geometric implementation of quantum
computation completely resilient to Markovian decoherence.
\end{abstract}

\pacs{03.65.-w 03.65.Vf 03.65.Yz}
 \maketitle

Adiabatic holonomies in quantum mechanics are unitary
transformations generated by slowly changing the Hamiltonian of a
quantum system through a set of parameters in a cyclic fashion.
The parameters define
a manifold and the holonomies depend only on the path followed by
the system in this space. In case the Hamiltonian is non-degenerate,
the holonomy is an abelian phase better known as Berry's
phase \cite{berry84a}. If the Hamiltonian is degenerate,
the holonomy is a non-abelian generalization of the phase
which induces transitions among states belonging to the degenerate
subspace \cite{wilczek84}.  Geometric phases, i.e., Berry's phase and its various generalizations,
were extensively studied in the
1980's \cite{shapere89a} and recently became, once more, a
fashionable subject since it was proposed that quantum computation
(QC) could be implemented in a fault-tolerant way using geometric
transformations \cite{jones99a,zanardi99}. The most general scheme
of this kind is called holonomic quantum computation (HQC) and it
involves constructing a universal set of quantum gates by making the system acquire a
succession of abelian and non-abelian geometric phases. Since
holonomies depend only on geometric properties of the parameter
space, HQC is robust, by construction, to any kind of error that leaves these
properties invariant. This is considered to be the strongest
advantage of the scheme.

In any realistic implementation of quantum computation
one must consider that, in practice, quantum systems are never isolated. The
interaction with the environment effects that initially pure quantum
states decay into statistical mixtures (called mixed states)
through a process known as decoherence \cite{deco}. It is known that this process
is a severe limiting factor for quantum
computation. Therefore, in a pragmatic realization of HQC schemes,
it is crucial to define holonomies for mixed states, and
understand how they are generated under the presence of
decoherence.
So far, non-abelian holonomies
have been only investigated for pure quantum states evolving under
unitary transformations. It is now necessary to investigate
to what extent the geometric properties
on which holonomies depend, are left invariant under quantum noise.

 Although claims that quantum gates of geometric
origin offer increased fault-tolerance can be found throughout
the literature, a comprehensive analysis on the robustness to errors
of these schemes has not yet been presented.
Partial results, however, have been obtained:
For both, abelian and non-abelian phases, some errors of
classical origin have been investigated \cite{class} and
the effect of quantum noise has been analyzed only for the abelian case in \cite{ericsson02,carollo03a,carollo03b}.
In particular, Ref. \cite{carollo03a} uses the quantum jump approach to study the
effects of decoherence in the adiabatic and non-adiabatic case. There the spin-1/2
Berry phase was found to be robust to
the depolarizing channel in the no-jump trajectory. Moreover, resilience to the phase flip
error is guaranteed for any number of jumps. In case the particle is driven by a
quantized field, Berry's phase is found to be robust to field decoherence to
second order \cite{carollo03b}.

The robustness of the abelian phase to a number of decoherence
effects suggests that non-abelian phases might behave similarly.
Motivated by this, in this Letter, we investigate the effects of
decoherence on non-abelian holonomies under the Markovian
assumption, where environmental memory effects are negligible. We
demonstrate the application of our general framework to the study
of decoherence in HQC. Independently of the physical
implementation considered, we estimate the errors, produced by
generating a universal set of holonomic gates in the presence of
an environment. We pinpoint to which specific types of errors the
scheme is fallible. These results constitute the basis of a
framework to perform universal HQC resilient to Markovian
decoherence. Decoherence can produce two kinds of errors: those
which take the system out of the degenerate subspace and those
which take place within the subspace. The first kind of error can
be eliminated by working in the ground state and considering a
system where the energy gap with the first excited state is very
large. Our analysis concerns the second kind of errors. We
consider only adiabatic holonomies because the non-adiabatic case
is not well understood. It is not clear how to separate, in
general, dynamic from geometric evolution in non-adiabatic
holonomies.

 Our work focuses on the applications
of holonomies in quantum computation, but geometric phases are a
fundamental aspect of quantum mechanics with applications in many
fields. They have manifestations \cite{shapere89a} that range from
low to high energy physics, appearing in optical and solid state
systems, in molecular and atomic physics, and are at the heart of
phenomena such as anomalies in gauge field theories, fractional
statistics, Aharonov-Bohm and quantum Hall effect. The discussion
presented here is relevant to any physical situation involving
holonomies where decoherence plays a role.


To introduce the concept of holonomy we consider a Hamiltonian
$H(\lambda_{0})=H_{0}$ describing an energy degenerate
N-dimensional quantum system. The Hilbert space has $k$ subspaces
with corresponding energy $E_{k}$. Our analysis will be restricted
to one of the degenerate subspaces with ${n}$-fold degeneracy and
energy $E$. The system is initially prepared in a state belonging
to this subspace $|\psi^{0}_{i}\rangle=|\psi_{i}(0)\rangle$ and
the Hamiltonian of the system varied through a set of parameters
${\lambda}$ on a control manifold $\mathcal{M}$ in an adiabatic
way. The adiabatic theorem states that if the Hamiltonian is
varied sufficiently slowly with respect to any time-scale
associated with the dynamics, the system will remain in a state
belonging to the subspace corresponding to energy $E$. In other
words, there is no level-crossing. The degeneracy structure of the
Hamiltonian is preserved throughout the evolution which we can
write as $H(\lambda)=V(\lambda)H_{0}V^{\dagger}(\lambda)$, where
$V(\lambda)$ is unitary. If the Hamiltonian returns to its initial
value after a time $T$, $H(T)=H_{0}$, describing a closed curve
$C$ in $\mathcal{M}$, the state $|\psi^{0}_{i}\rangle$ is mapped
to $e^{-i E T}{U}_{C}(\lambda)|\psi^{0}_{i}\rangle$. The
transformation ${U}_{C}(\lambda)$ is called the holonomy and in
the following, we will derive it.

By changing the Hamiltonian through the set of parameters
$\lambda$, the state of the system is parallel transported in
parameter space. A rule for parallel transport in the manifold,
i.e. a connection, must be specified since there is no unique way
of parallel transporting a vector. A connection is provided by
requiring that the state $|\psi_{\alpha}(\lambda)\rangle$ remains
normalized through parallel transport,
$\langle\psi_{\beta}|\frac{\partial}{\partial\lambda_{\mu}}|\psi_{\alpha}\rangle
= 0$.

In terms of the local reference basis of the degenerate subspace
$\{\ket{\phi_{\alpha}(\lambda)}\}$, the state at any point of the
adiabatic path in parameter space is expressed as, $
\ket{\psi_{\alpha}(\lambda)}=U_{\alpha\beta}(\lambda)\ket{\phi_{\beta}(\lambda)}$,
where $U$ is unitary and $\ket{\psi_{\alpha}(\lambda)}$
corresponds to the solution of the Schr\"{o}dinger equation for
the initial condition
$\ket{\psi^{o}_{\alpha}}=\ket{\phi_{\alpha}(0)}$. The parallel
transport condition then reads,
\begin{equation}\label{parallel}
{U}^{\dagger}_{\gamma\delta} \frac{\partial
U_{\alpha\beta}}{\partial\lambda_{\mu}} P_{\beta \gamma} +
U^{\dagger}_{\gamma\delta}U_{\alpha\beta}A_{\beta\gamma}=0,
\end{equation}
with $P_{\beta\gamma}= \langle\phi_{\gamma}|\phi_{\beta}\rangle$
and $A_{\beta\gamma}=\langle\phi_{\gamma}|\frac{\partial}
{\partial\lambda_{\mu}}|\phi_{\beta}\rangle$, in the Wilczek-Zee
\cite{wilczek84} notations. The matrix $P$ is hermitian and in the
case when the states of the local basis are orthogonal it is equal
to the identity. In the presence of decoherence, the latter is not
true in general. We ignored the constant term $EP_{\beta\gamma}$
which produces the global dynamical phase $e^{-i E T}$. Solving
for $U$ when $P$ is invertible, $U^{-1}\dot{U}=-AP^{-1}$, and
integrating over the closed path $C$ we obtain the holonomy,
\begin{equation}\label{eq:holonomy}
U_{C}(\lambda)={\bf P}e^{{-\int_{C}AP^{-1}}}.
\end{equation}
where ${\bf P}$ is the path ordering operator. In terms of the
eigenstates of the initial Hamiltonian $H_{0}$, we have
$A_{\beta\gamma}=\langle\psi^{0}_{\gamma}|V^{\dagger}\dot{V}|\psi^{0}_{\beta}\rangle$.
The dimensionality $n$ of the holonomy equals the degree of
degeneracy of the eigenspace. Berry phase is the special case when
the eigenspace is non-degenerate, and the unitary transformation
is then one dimensional, i.e. a complex number. The holonomy
(\ref{eq:holonomy}) depends only on the path followed in parameter
space, and transforms, under gauge transformation $g$, as
$U_{C}(A)\rightarrow gU_{C}(A) g^{-1}$ \cite{wilczek84}. We point
out that all previous considerations can be extended to the case
of an open curve without any modifications since gauge invariance
is not relevant in our context.

To investigate the effects of decoherence in the non-abelian
geometric evolution of states described above, we employ the
quantum jump approach. The master equation ($\hbar=1$)
\begin{equation}
  \label{eq:mastereq}
  \dot{\rho}=\frac{1}{i}[H(\lambda),\rho]-\frac{1}{2}\sum_{k=1}^{n}
  \{L_k^\dagger L_k\rho+\rho L_k^\dagger L_k\ - 2 L_k
  \rho L_k^\dagger\}
\end{equation}
dictates, in the Markovian approximation, the evolution of the
system described by the density operator $\rho$. The commutator
generates, through the Hamiltonian $H(\lambda)$, the coherent part
of the evolution and the second part represents the effect of the
environment on the dynamics of system. The operators $L_k$ are
called Lindbladian, and by prescribing them, one models different
decohering processes. Equation (\ref{eq:mastereq}) is in general
very difficult to solve but the quantum jump approach provides an
ingenious solution to this problem. Consider that the time
evolution of the density matrix, for small time intervals $\Delta
t$, can be written as
\begin{equation}
  \label{eq:decomp}
  \rho(t+\Delta t)\approx \sum_{k=0}^{n} W_k \rho(t) W_k^\dagger,
\end{equation}
where the operators $W_k$ are complete positive maps fulfilling
the completeness relation $\sum_{k=0}^nW_k^\dag W_k=\one$. By
setting $W_0=\one-i\tilde{H}\Delta t$ and $W_k=\sqrt{\Delta t}
L_k$ with $\tilde{H}$ a non-Hermitian effective Hamiltonian,
\begin{equation}
  \label{eq:nonHerHam}
  \tilde{H}=H-\frac{i}{2}\sum_{k=1}^n L_k^\dagger L_k,
\end{equation}
the dynamics of the system is approximated by dividing the total
evolution time $T$ into a sequence of discrete intervals $\Delta t
= \frac{T}{N}$. $W_0$ and $W_k$ are called the "no-jump" and jump
operators respectively. According to Eq.~(\ref{eq:decomp}), the
state of the system, after any time step $t_m=m\Delta t$, evolves
into $\rho(t_{m+1})=W_k \rho(t_m) W_k^\dagger$ (up to first order
in $\Delta t$), with probability $p_k=Tr{W_k\rho(t_m)
W_k^\dagger}$. The time evolution of the system is then
calculated for a set of possible trajectories containing, each one
of them, different numbers of jumps occurring at different times.
Each trajectory is defined as a chain of states obtained by the
action of a sequence of operators $W_k$ on the initial state. For
example, for an initial pure state $\ket{\psi_0}$, the
(non-normalized) state of the system, after the $m$-th step along
the $i$-th trajectory, is given by:
\begin{equation}
\ket{\psi^{(i)}_m}=\prod_{l=1}^{m}W_{i(l)}\ket{\psi_0},
\end{equation}
where $i(l)$ stands for the $l$-th element of a sequence of indexes
with values from ${0\dots n}$. Each trajectory is
represented by a discrete sequence of pure states
$\{\psi_{0},\psi^{(i)}_{0},\dots,\psi^{(i)}_{N}\}$. The dynamics
given by the master equation is recovered by summing incoherently
all the states associated to each trajectory, and taking the
continuous limit $\Delta t \to 0$.

The no-jump trajectory corresponds to the case in which no decay
occurs. The evolution of a quantum state along this trajectory is
obtained by the repeated action of the operator $W_0$ and taking
the continuous limit $N\to \infty$. This yields a dynamics
governed by the complex effective Hamiltonian $\tilde{H}$:
\begin{equation}
\label{eq:nojumpcont} i\frac{d}{dt}
\ket{\psi(t)}=\tilde{H}\ket{\psi(t)},\quad
\ket{\psi(0)}=\ket{\psi_0}.
\end{equation}
Since the Hamiltonian is non-hermitian, the corresponding
eigenstates are non-orthogonal.

We are now ready to consider the case in which a non-abelian
holonomy is generated by the transformation
$H(\lambda)=V(\lambda)H_{0}V^{\dagger}(\lambda)$ when the system's
dynamics governed by (\ref{eq:mastereq}). Note that it is only
possible to generate the holonomy when the interaction of the
system with the environment is such that the degeneracy structure
of the Hamiltonian is preserved. This is the case when
$\kappa=\sum_{k=1}^n L_k^\dagger L_k=\alpha\one$ or $\kappa=\alpha
H$. The holonomy in the no-jump trajectory is then the same as the
one acquired by an isolated system evolving under the same
Hamiltonian $H$. This is because the eigenstates of $\tilde{H}$
coincide with the eigenstates of $H$, so that $P$ is proportional
to the identity and  the connection in the no-jump trajectory
$(AP^{-1})^{(0)}$ is equal to $A=V^{\dagger}\dot{V}$. The
interaction with the environment only produces an overall
visibility factor $e^{\alpha t/2 }$ for $\kappa=\alpha\one$ and
$e^{\alpha Et/2}$ for $\kappa=\alpha H$. The factor is small for
low decoherence rates and in the second case, it is eliminated by
working in the ground state. In other words, for sources of
decoherence that satisfy these conditions, the holonomy is robust
in the no-jump trajectory.

Let us consider a more interesting case, the trajectory in which there is one jump $W_{i}$ at $\lambda_1$. During the
adiabatic evolution, the system evolves under a no-jump trajectory from $\lambda_{0}$ to $\lambda_{1}$, then the jump
occurs instantaneously and the system continues to evolve by the transformation $V(\lambda)$ until time T,
corresponding to $\lambda_{f}$. Using the composition rule for holonomies, $U_{1}=({\bf P}e^{-\Gamma_{1}})({\bf P}
e^{-\Gamma_{0}})$ where
\begin{eqnarray}
\Gamma_{0}&=&
{\int_{0}^{\lambda_{1}}(AP^{-1})^{(0)}}={\int_{0}^{\lambda_{1}}A},\\
\Gamma_{1}&=&{\int_{\lambda_{1}}^{\lambda_{f}}{(AP^{-1})}^{(1)}}.
\end{eqnarray}
Since the one-jump connection for
$W^{\dagger}_{i}W_{i}=\alpha_{i}\one$ is given by,
\begin{eqnarray}
(AP^{-1})^{(1)}_{\beta\gamma}&=&\langle\phi^{\prime}_{\delta}|V^{\dagger}(\lambda)\dot{V}(\lambda)|\phi^{\prime}_{\beta}\rangle
P^{-1}_{\delta \gamma},\\
|\phi^{\prime}_{\gamma}\rangle&=&W_{i}e^{\Gamma_{o}}|\phi^{0}_{\gamma}\rangle, \\
P_{\beta\gamma}&=& \langle\phi^{\prime}_{\gamma}|\phi^{\prime}_{\beta}\rangle = \alpha_i \delta_{\beta\gamma},
\end{eqnarray}
the holonomy after the jump is
\begin{equation} \label{eq:jump}
{\bf P}e^{-\Gamma_{1}}= {\bf P}
e^{-\frac{1}{\alpha_i}\int_{\lambda_{1}}^{\lambda_{f}} W_i^\dagger
AW_i}.
\end{equation}So in the
one-jump trajectory $U_{1}={\bf
P}e^{-\frac{1}{\alpha_{i}}W_{i}\Gamma_{0}^{\prime}W_{i}}{\bf P
}e^{-\Gamma_{0}}$. This result can then be generalized to a
trajectory for which n jumps occur
\begin{eqnarray}U^{i}_{n}&=&\prod_{l=1}^{n}
{\bf P} e^{-\frac{1}{\alpha_{i}}W^{\dagger}_{i(l)}\Gamma_{0}^{i(l)}W_{i(l)}}{\bf P}e^{-\Gamma_{0}}\\
\Gamma_{0}^{i(l)}&=&\int_{\lambda_{l}}^{\lambda_{l+1}}A.
\end{eqnarray}
Note that we can write,
\begin{equation} \label{eq:jump}
{\bf P}
e^{-\frac{1}{\alpha_{i}}W^{\dagger}_{i(l)}\Gamma_{0}^{i(l)}W_{i(l)}}=\frac{1}{\alpha_i}
W_{i(l)}^\dagger \left({\bf P}
e^{-\int_{\lambda_{1}}^{\lambda_{f}} A}\right) W_{i(l)} .
\end{equation}
We now consider the application of our results in QC, where the
idea of using quantum systems to perform computations more
efficiently than classical computers is investigated
\cite{nielsen00}. Geometric phases play an important role in this
field since quantum gates can be implemented in a geometric way.
Jones {\it et. al.} \cite{jones99a} proposed a scheme for
generating quantum single and two-qubit phase gates using Berry's
phase.

The generalized scheme to implement QC by geometric means is HQC
\cite{zanardi99}. In this scheme the input information is encoded
in the $n$ base states of a given degenerate subspace at
$H(\lambda_{0})$. By renormalizing one can choose the energy of
the subspace $E=0$. The gates are generated by the holonomic
evolutions described above (this includes abelian and non-abelian
phases) which are produced by slowly changing the initial
Hamiltonian through the set of parameters $\lambda$. After
$\lambda$ completes a loop $C$ in $\mathcal{M}$ rooted at
$\lambda_{0}$, the initially prepared state $|\psi_{in}\rangle$ in
which the information was encoded is mapped to an output state
$|\psi_{out}\rangle={U}_{C}(\lambda)|\psi_{in}\rangle$ where
${U}_{C}(\lambda)$ is the quantum gate. To perform a given gate it
is necessary to find the path or succession of paths in parameter
space which yields such a gate. A quantum algorithm is built from a
sequence of gates such that the final state corresponds to the
solution of the computational problem. Any unitary transformation,
thus any algorithm, can be approximated by a strategical
succession of closed paths in the control manifold $\mathcal{M}$.
As long as the adiabatic condition holds, the computation does not
depend on the rate at which the control loops are driven. The
gates performed in such a way depend only on geometric properties of
the parameter space. Thus, any computational error which is
path-preserving does not change the transformation. Hence, errors
such as fluctuations in the driving parameters and systematic
errors are automatically avoided. However, the environment might induce some
fluctuations which are not path-preserving; it is then important
to quantify them.

Investigating the effects of decoherence in HQC becomes crucial
for the scheme to be realizable. Our quantum jump formalism can be
used to investigate decoherence in a universal set of gates
generated by holonomies. By universal we mean that any unitary
evolution can be constructed from this set of elementary gates
\cite{uni}. The set consists of two 1-qubit gates of the form
$U_{i}=e^{i\theta_{i}\sigma_i}$ where $\sigma_{i}$ are Pauli
matrices, and one 2-qubit gate
$U_{3}=e^{i\phi\sigma_i\otimes\sigma_j}$. For example,
$U_{1}=e^{i\theta_{1}\sigma_1}$, $U_{2}=e^{i\theta_{2}\sigma_2}$
and $U_{3}=e^{i\phi\sigma_1\otimes\sigma_1}$ are universal since
any transformation in $SU(4)$ can be generated from them. Consider
that these gates are performed by holonomic evolution as described
in \cite{holgat} but the system interacts with the environment.
Errors within the subspace for one qubit gates are proportional to
$\sigma_{-}$, $\sigma_{+}$ and $\sigma_{i}$, with $i=1,2,3$. For
$L_i=\sqrt{\alpha}\sigma_{i}$ and $\alpha$ the decoherence rate of
the error, we obtain $L^{\dagger}_{i}L_{i}=\alpha\one$. Thus the
universal set of gates is robust to these errors in the no-jump
trajectory. When there is a jump in the trajectory, it is easy to
see from eq. (\ref{eq:jump}) that the gate $U_{i}$ is robust to
errors $L_{j}$ when $i=j$ and for errors with $j\neq i$ one
obtains a change in sign (since
$\sigma^{\dagger}_{j}\sigma_{i}\sigma_{j}=-\sigma_{i}$). For
example, if one jump at occurs at $\lambda_{1}$, the gate $U_{i}$
becomes $U_{i}=\exp i(\theta_{1}-\theta_{2})\sigma_{i}$ where
\begin{eqnarray}
\theta_{1}=\int_{\lambda_{0}}^{\lambda_{1}}\Omega,\quad
\theta_{2}=\int_{\lambda_{1}}^{\lambda_{f}}\Omega.
\end{eqnarray}
Here $\Omega$ corresponds to the solid angle subtended by the
evolution of the system in parameter space. The angle
$\theta=\theta_{1}+\theta_{2}$ corresponds to the gate without
errors. Generalizing to n errors of the type $L_{j}$ we find that
$U_{i}(n)=e^{i\theta_{e}\sigma_{i}}$ with
$\theta_{e}=\sum_{m}(-1)^{m}\theta_{m}$ and
$\theta_{m}=\int_{\lambda_{m}}^{\lambda^{m+1}}\Omega$. When errors
$\sigma_{\pm}$ occur, the degeneracy structure of the Hamiltonian
is preserved only for initial Hamiltonians which are proportional
to $\sigma_{3}$ since
$\sigma_{\pm}\sigma_{\mp}=\frac{1}{2}(\one\pm\sigma_{3})$. In this
case the holonomy is robust in the no-jump trajectory. For a
single jump $\sigma_{\pm}$, $P$ is not invertible but we can
calculate the gates from eq. \ref{parallel}. Unfortunately, the
gates are completely lost, $e^{i\theta\sigma_{i}}$ becomes a
$U(1)$ phase after one jump. This is the main source of error that
must be corrected.

The analysis of the two-qubit gates follows directly from our
previous conclusions. As an example we consider errors of the form
$\sigma_i\otimes\sigma_j$. For simplicity let us consider the gate
$U_{3}=e^{i\phi\sigma_1\otimes\sigma_1}$, any other gate can be
analyzed in the same way. The gate is robust to any number of
jumps of the type $\sigma_1\otimes\sigma_1$,
$\sigma_1\otimes\one$, $\one\otimes\sigma_1$ and
$\sigma_2\otimes\sigma_2$ but changes in signs occur for
$\one\otimes\sigma_2$, $\sigma_2\otimes\one$,
$\sigma_1\otimes\sigma_2$ and $\sigma_2\otimes\sigma_1$.
Nevertheless in general the most common non local error is
$\sigma_i\otimes \sigma_i$, to which the 2-qubits gate is robust.

We have presented a general scheme to study the effects of
Markovian decoherence in the generation of non-abelian adiabatic
holonomies. We applied it to analyze the effects of the
environment on a universal set of holonomic quantum gates. A
scheme for quantum computation completely robust to Markovian
decoherence can be constructed by using HQC and finding a way to
correct for the errors pointed out in this Letter by other means.
This could possibly be achieved using error correction techniques.
An example of the use of geometric phases to produce an error
correcting code is explored in \cite{ecorrect}. We are currently
working on the details of such a scheme. Since holonomies are a
main ingredient in loop quantum gravity it is likely our scheme
can be applied to study decoherence in this field.

We would like to thank Frederic P Schuller for his valuable
comments.


\end{document}